\documentclass[aps,prl,twocolumn,superscriptaddress,longbibliography]{revtex4-2}

\usepackage[T1]{fontenc}
\usepackage[utf8]{inputenc}
\usepackage{amsmath,amssymb,bm}
\usepackage{graphicx}
\graphicspath{{./figs/}}
\usepackage{xcolor}
\usepackage[colorlinks=true,linkcolor=blue,citecolor=blue,urlcolor=blue]{hyperref}
\usepackage[nameinlink,capitalise,noabbrev]{cleveref}
\usepackage[normalem]{ulem}
\usepackage{pdfpages}

\definecolor{myRGB}{RGB}{195,26,90}

\newcommand{\TP}[2]{#2}

\begin{document}
\makeatletter
\let\LS@rot\relax
\makeatother
\date{\today}
\title{Penetration Resistance Force in Granular Matter}
\author{Tongge Wen}
\thanks{These authors contributed equally to this work.}
\affiliation{Department of Engineering Mechanics, State Key Laboratory for Strength and Vibration of Mechanical Structures, School of Aerospace Engineering, Xi'an Jiaotong University, Xi'an 710049, China}
\affiliation{School of Automation, Beijing Institute of Technology, Beijing 100081, China}
\author{Xiaoyu Yang}
\thanks{These authors contributed equally to this work.}
\affiliation{School of Automation, Beijing Institute of Technology, Beijing 100081, China}
\author{Guang-Kui Xu}
\affiliation{Department of Engineering Mechanics, State Key Laboratory for Strength and Vibration of Mechanical Structures, School of Aerospace Engineering, Xi'an Jiaotong University, Xi'an 710049, China}
\author{Thorsten P\"oschel}
\affiliation{Institute for Multiscale Simulation, Friedrich-Alexander-Universit\"at Erlangen-N\"urnberg, Cauerstra{\ss}e 3, 91058 Erlangen, Germany}
\author{Xiangyuan Zeng}
\email{Corresponding author: zeng@bit.edu.cn.}
\affiliation{School of Automation, Beijing Institute of Technology, Beijing 100081, China}

\begin{abstract}
Penetration resistance of an intruder into granular media is commonly described as a friction-controlled response in which particle-scale shape effects enter only through an effective macroscopic friction coefficient. 
\TP{1}{Here we show that enhanced penetration resistance in packings
of elongated particles is coupled to intrusion-induced alignment and anisotropic stress transmission.} 
Discrete element simulations reveal that penetration resistance increases with the particle aspect ratio and collapses onto a common scaling law governed jointly by friction and particle shape. The increase of resistance 
occurs despite a progressive reduction of the stagnant zone beneath the intruder, indicating that resistance cannot be explained solely by the amount of displaced material. Instead, the
process of penetration induces an orientational self-organization of elongated particles, forming a highly anisotropic load-bearing force network that modifies stress transmission. We identify a coupled mechanism in which particle elongation promotes stress-induced orientational organization, whereas interparticle friction stabilizes the resulting force network against sliding. This coupling mechanism is the microscopic origin of the enhanced penetration resistance.
\end{abstract}
\maketitle

The quasi-static penetration of a rigid intruder into a granular medium is a canonical problem that bridges fundamental non-equilibrium physics and diverse applications, ranging from planetary exploration to robotic locomotion \cite{Askari:2016, Roth:2021a, Morota:2020}. The penetration resistance arises from stress transmission through the network of frictional contacts \cite{Irmer:2025, Askari:2016, Kim:2020, LopezRodriguez:2017, Clark:2012}. Under quasistatic conditions, experiments and simulations consistently show that, after a short initial transient, the penetration resistance increases approximately linearly with penetration depth \cite{Wen:2026,Lin:2024,Zhang:2022,Roth:2021b}. 

Frictional hydrostatic \cite{Brzinski:2013} and modified Archimedes-type models \cite{Kang:2018} obtain the resistance by integrating depth-dependent frictional stresses over the intruder surface, leading to a penetration force-depth slope $K$ that scales with the bulk effective friction coefficient, $\mu_\text{eff}$ \cite{Seguin:2011, DiazMelian:2020, Zheng:2018, Pravin:2021, LopezRodriguez:2017, Yin:2025, Feng:2019, PicaCiamarra:2004}. 
\TP{1}{In these friction-based models, particle-scale shape effects are represented through the effective bulk friction coefficient $\mu_\text{eff}$. Whether this coefficient alone can account for the shape dependence of penetration resistance in packings of elongated particles remains an open question.}

\TP{1}{Previous studies have examined how particle alignment affects bulk packing properties and macroscopic friction \cite{Fan:2024,Bilotto:2025,Zou:2023,Rahim:2024}.
Incorporating orientational order into kinetic theory has also enabled predictions for homogeneous shear flows of frictionless nonspherical particles \cite{Vescovi:2024}. 
In dense shear flows, collective alignment has been shown to control particle rotation \cite{Pol:2026}, while experiments have revealed a close correlation between particle orientation and rotational dynamics \cite{Pol2022}. Particle angular motion has also been linked to wall-friction mobilization in confined flows \cite{Pol2023}. 
These studies highlight the relevance of orientational order and rotational dynamics as ingredients in continuum descriptions of anisotropic granular flows. 
For penetration into dense, frictional packings of elongated particles, the central question is how the spatially heterogeneous orientational organization induced by the intruder affects stress transmission and penetration resistance.}

\begin{figure}[htbp]
\centering
\includegraphics[width=\columnwidth]{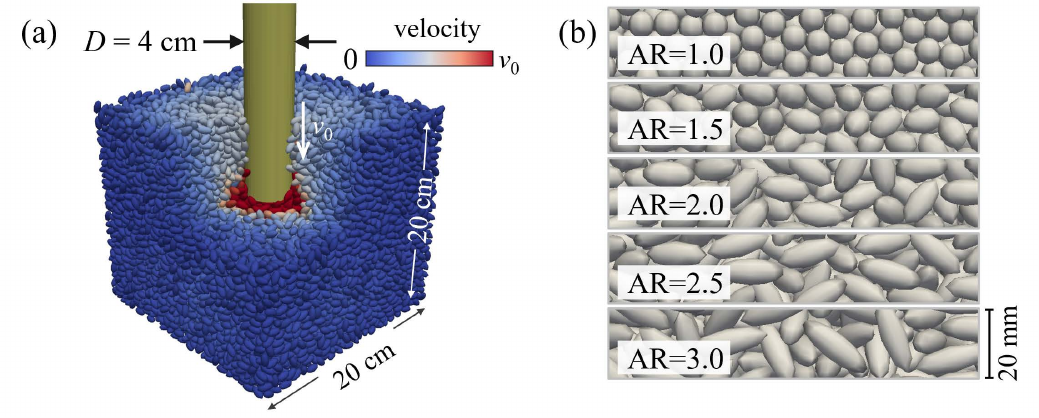}
\caption{a) Simulation setup; particles are colored by velocity magnitude. b) Particles with different aspect ratios ($\mathrm{AR}$).}
\label{fig:setup}
\end{figure}

To address this question, we perform discrete element simulations of quasistatic penetration. As shown in Fig.~\ref{fig:setup}a, a cylindrical intruder of diameter $D = 4\;\mathrm{cm}$ penetrates a granular bed at a constant velocity of $5\,\mathrm{mm/s}$. The simulation domain has a side length of $20\,\mathrm{cm}$ which is sufficiently large to minimize boundary effects \cite{Seguin:2008}. Elongated particles are modeled as multi-sphere particles \cite{Poeschel.1993,Poeschel.1993,Kruggel:2008,Favier:1999,Buchele.2026}, with aspect ratios ranging from $\mathrm{AR} = 1$ (spheres) to $\mathrm{AR} = 3$. All particles have the same minor semi-axis, $r = 3\,\mathrm{mm}$, such that particle shape is varied exclusively through the aspect ratio, as illustrated in Fig.~\ref{fig:setup}b. 
\TP{2}{All elongated particles with aspect ratios $\mathrm{AR}=1.5$ to $3$ are assigned the same microscopic Young's modulus, $E=0.2$ MPa. The microscopic Young's modulus of the spherical particles is calibrated to $E=0.12$ MPa to ensure comparable initial macroscopic compressive moduli across all particle shapes.}
The interparticle friction coefficient is varied independently, $\mu=0.3\dots 0.9$. Particle material density and gravitational acceleration are $\rho=1.2\,\mathrm{g/cm}^3$ and $g=9.81\,\mathrm{m/s}^2$, respectively. Details of the numerical implementation, calibration process, and validation experiments showing particle alignment during penetration are provided in the Supplemental Material \cite{SuppData} (see also references \cite{kruggel2008study,Buchele.2026_a, Buchele.2026_b, Moradian.2026_a, Moradian.2026_b,bshp:996,poschel1993static,Schwager2008,Mller2011} therein).

\begin{figure}[htbp]
\centering
\includegraphics[width=\columnwidth]{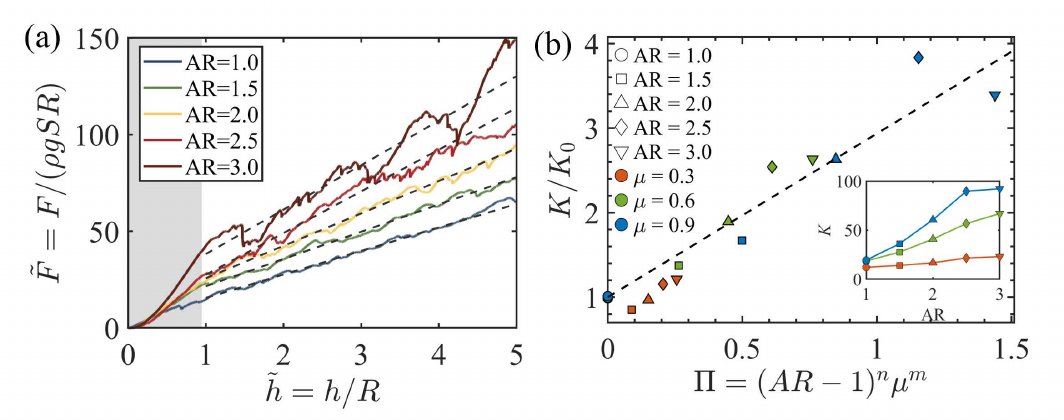}
\caption{a) Normalized penetration force--depth curves for different ARs. b) $K/K_0$ as a function of $\Pi$. Symbols denote simulation data, the dashed line shows the scaling curve. The inset shows the unscaled function $K(\mathrm{AR})$.}
\label{fig:setupA}
\end{figure}

The penetration resistance, $F$, acting on the intruder is measured as a function of penetration depth $h$, and reported in normalized form as $\widetilde{F}\equiv F/(\rho g S R)$ versus $\widetilde{h}\equiv h/R$, where $R$ is the intruder radius and $S=\pi R^2$ its projected area. As shown in Fig.~\ref{fig:setupA}(a), all force--depth curves exhibit a short nonlinear transient followed by a linear regime. While the overall response remains qualitatively similar across particle shapes, the slope of the linear regime increases systematically with $\mathrm{AR}$, indicating a higher penetration resistance in packings of elongated particles. We characterize the steady state response by the dimensionless resistance coefficient
\begin{equation}
    K\equiv \frac{\mathrm{d}\widetilde{F}}{\mathrm{d}\widetilde{h}}\,,
\end{equation}
defined as the slope of the linear regime. 

\TP{3}{Friction-based penetration models predict $K$ is proportional to the macroscopic effective friction coefficient $\mu_{\mathrm{eff}}$ \cite{Brzinski:2013,Cheng:2023}. Accordingly, we define a friction-based reference slope as $K_0=a\mu_\mathrm{eff}$, where $a$ is a dimensionless proportionality constant. 
In this study, $\mu_{\mathrm{eff}}$ is determined separately for each combination of $\mathrm{AR}$ and $\mu$ from the angle of repose of the corresponding granular assembly (see Supplemental Material \cite{SuppData}).}
\TP{3,5}{For spherical particles ($\mathrm{AR}=1$), the obtained slope $K$ coincides with the reference $K_0$. We therefore determine $a$ by fitting $K_0=a\mu_\mathrm{eff}$ to the spherical particle data and hold it fixed for all aspect ratios.}

For elongated particles, $\mathrm{AR}>1$, the resistance systematically exceeds the \TP{1}{friction-based reference}, $K>K_0$ (Fig.~\ref{fig:setupA}a). As shown below, the process of penetration induces local particle alignment, which restructures the force network. \TP{1}{A scalar bulk friction coefficient does not explicitly resolve this spatially heterogeneous orientational organization.}

\TP{3}{To quantify the resistance enhancement beyond the friction-based reference, we define the relative resistance enhancement $\Delta_K\equiv(K-K_0)/K_0$. This enhancement vanishes in the spherical limit $\mathrm{AR}\to1$, motivating the inclusion of the factor $(\mathrm{AR}-1)^n$. Although $\mu_{\mathrm{eff}}$ demonstrates the combined influence of $\mathrm{AR}$ and $\mu$ on the bulk frictional response, it does not address how elongated particles reorient locally during penetration and how this reorientation influences the force network. Therefore, the influence of microscopic friction on stress transmission through this evolving force network may not be fully captured by $\mu_{\mathrm{eff}}$. We therefore preserve an explicit dependence on $\mu$ in the scaling relation and introduce the dimensionless parameter $\Pi\equiv(\mathrm{AR}-1)^n\mu^m$. We adopt a linear dependence of $\Delta_K$ on $\Pi$, yielding}
\begin{equation} 
K=K_0(1+b\Pi)\,. 
\label{eq:scaling} 
\end{equation}

Plotting $K/K_0$ against $\Pi$ reveals a collapse across all simulated conditions (Fig.~\ref{fig:setupA}b), supporting the proposed scaling. 
\TP{5}{The prefactor $a$ is determined independently from the $\mathrm{AR}=1$ data, yielding $a=78.558 \pm 0.557$. With $a$ fixed, a fit to the data across all aspect ratios yields $b=1.935 \pm 0.203$, $n=0.764 \pm 0.162$, and $m=1.571 \pm 0.266$, with a coefficient of determination of 0.947.}
Importantly, the independent datasets for $\mathrm{AR}=1.2$ and $\mathrm{AR}=2.2$ also collapse onto the master curve without adjusting the exponents (see Supplemental Material \cite{SuppData}), demonstrating that Fig.~\ref{eq:scaling} captures a physically meaningful scaling rather than merely fitting the data. 


\begin{figure}[htbp]
\centering
\includegraphics[width=\columnwidth]{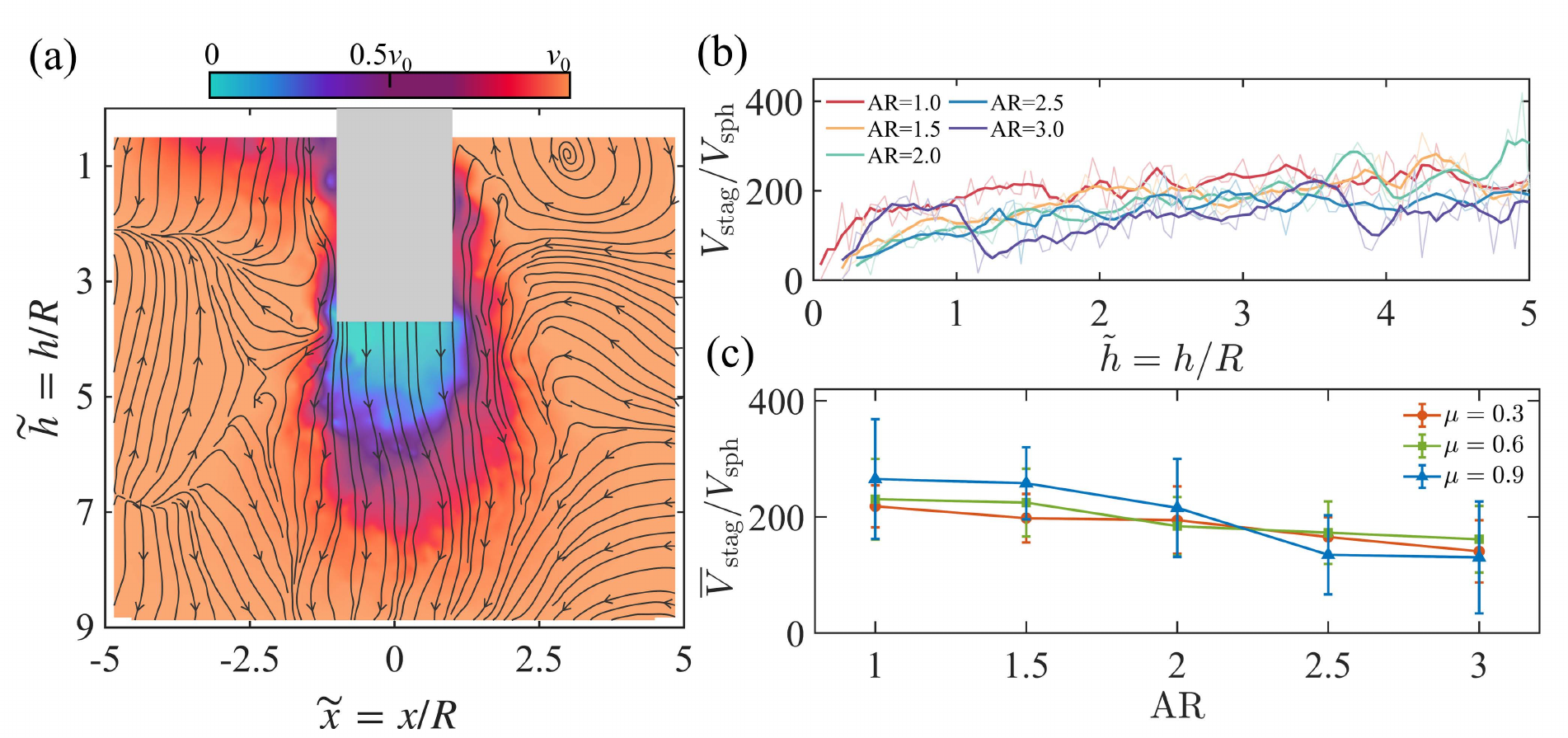}
\caption{\TP{6}{Stagnant zone formed ahead of the intruder. a) Schematic illustration of the stagnant zone. b) Evolution of the normalized stagnant zone volume $V_{\mathrm{stag}}/V_{\mathrm{sph}}$, with normalized penetration depth  $\tilde{h}=h/R$ for different aspect ratios at $\mu=0.3$. Thin, light-colored lines show the raw data, while thick, dark-colored lines show the data smoothed using a moving average over $\Delta\tilde{h}=0.25$. c) Time-averaged normalized stagnant zone volume ${\bar V}_{\mathrm{stag}}/V_{\mathrm{sph}}$ as a function of aspect ratio for three friction coefficients. The time average is calculated over 4 to 20 s, corresponding to $1\leq\tilde{h}\leq5$. Error bars indicate temporal standard deviations. In both b) and c), the stagnant zone volume is normalized by $V_{\mathrm{sph}}=4\pi r^3/3$, where $r=3$ mm is the particle short radius. The stagnant zone identification and the calculations of its volume and particle number are described in the Supplemental Material \cite{SuppData}.}}
\label{fig:stagnant}
\end{figure}

\TP{6}{Higher penetration resistance has often been associated with an expansion of the load-bearing region ahead of the intruder, involving a larger number of load-bearing particles \cite{Cheng:2023,Kang:2018}. We therefore examine how the spatial region of the stagnant zone changes with particle shape. The stagnant zone is defined kinematically as the region comprising particles that move approximately at the same velocity as the intruder, with negligible relative motion.
Figure~\ref{fig:stagnant}b shows the evolution of the stagnant zone volume for different aspect ratios at $\mu=0.3$. Although the volume fluctuates during penetration, its time-averaged value provides a quantitative measure for comparing different particle shapes. We calculate this average over 4 to 20 s, corresponding to $1\leq\tilde{h}\leq5$ in the steady penetration regime. As shown in Fig.~\ref{fig:stagnant}c, the average stagnant zone volume decreases systematically with increasing $\mathrm{AR}$ at all three friction coefficients, whereas the penetration resistance increases.} 
Hence, the enhanced resistance is not accompanied by an expansion of the stagnant zone.

\begin{figure}[htbp]
\centering
\includegraphics[width=\columnwidth]{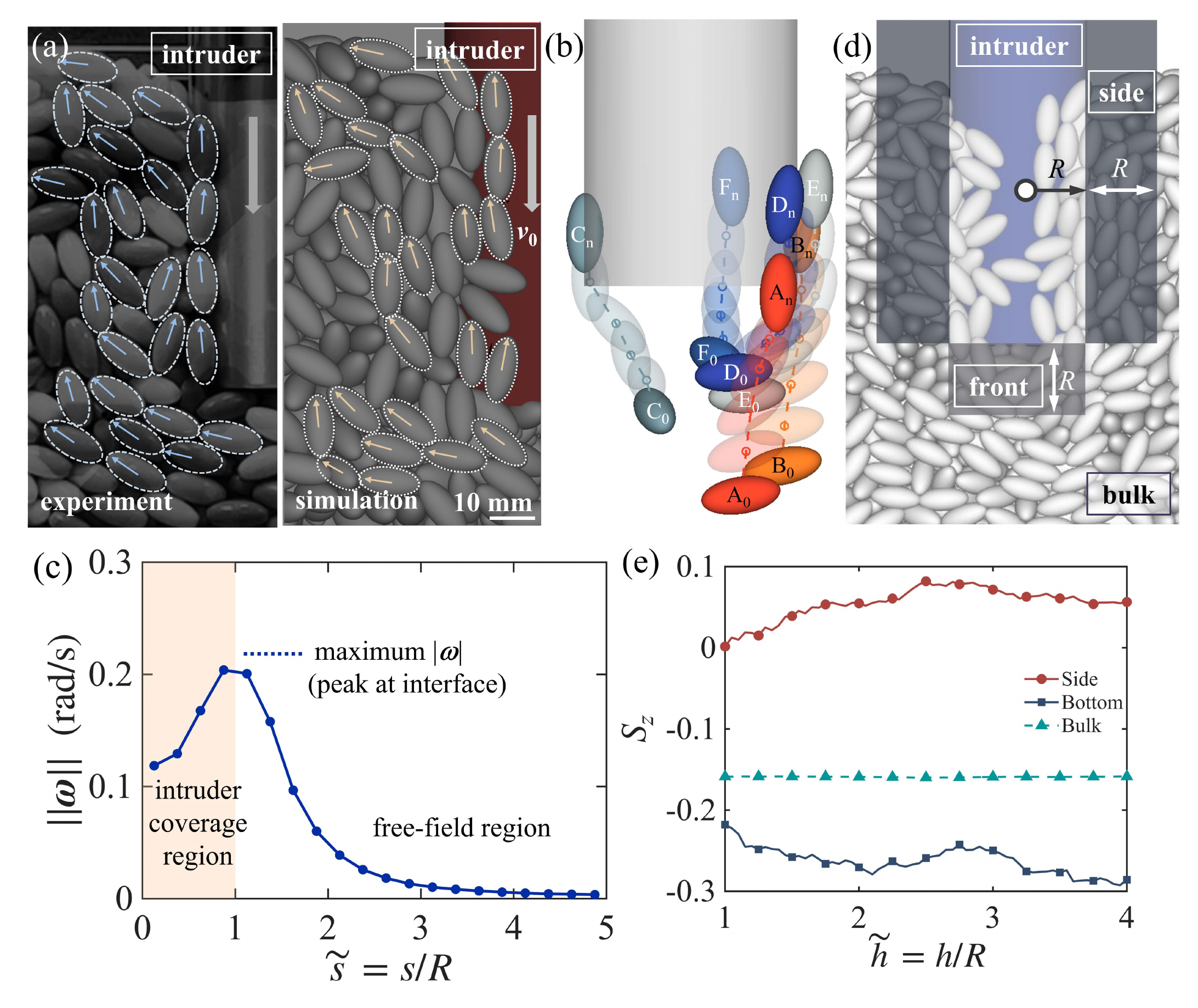}
\caption{a) Particle orientations near the intruder in experiments and simulations. The experimental setup and image analysis procedure are described in the Supplemental Material \cite{SuppData}.
b) Trajectories and orientation evolution of five representative particles during penetration.
\TP{7}{c) Radial profile of the mean particle angular velocity magnitude as a function of the normalized radial distance $\tilde{s}$. The statistics are calculated using the particles over a vertical interval extending $2R$ above and below the intruder bottom surface, where $R$ is the intruder radius.}
d) Partition of the granular bed into the compression-dominated front region, the shear-dominated side region, and the undisturbed bulk region. The regions are defined relative to the instantaneous intruder position.
e) Evolution of the orientational order parameter $S_z$ in the front, side, and bulk regions. The front region retains a fixed shape relative to the intruder, while the side region extends along the penetration path as the intruder advances.
\TP{7}{Sensitivity analyses of the radial bin width and vertical analysis region used for the angular velocity profiles, as well as the spatial regions used to calculate $S_z$, are provided in the Supplemental Material \cite{SuppData}.}
}
\label{fig:orientation}
\end{figure}

This finding motivates us to examine the orientational reorganization of elongated particles around the intruder.
In the compressed front region, particles preferentially align perpendicular to the penetration direction. In the side region, their long axes align with the local shear direction (see also Supplementary Videos). This behavior is observed in both experiments and particle simulations (Fig.~\ref{fig:orientation}a). 
Figure~\ref{fig:orientation}b tracks five representative particles, illustrating how their orientations evolve as they migrate relative to the intruder from the compressed front into the shear-dominated side region, transitioning from transverse to flow-aligned.
\TP{7}{Figure~\ref{fig:orientation}c quantifies particle rotation through the radial profile of the mean angular velocity magnitude, $||\boldsymbol{\omega}||=(\omega_x^2+\omega_y^2+\omega_z^2)^{1/2}$. 
Particles are assigned to radial bins according to their instantaneous positions relative to the intruder axis, with an output interval of 0.4 s from 4 to 20 s.} 
The resulting profile is nonmonotonic, reaching a maximum near the interface between the compression-dominated front region and the shear-dominated side region. Beyond this transition, the angular velocity magnitude decreases rapidly with radial distance. This indicates that particle rotation is concentrated near the intruder, particularly in the transition region where compressive and shear deformation coexist, and weakens rapidly farther away.

When we partition the granular volume relative to the instantaneous intruder position into a compression-dominated front region, a shear-dominated annular side region, and an undisturbed bulk region (Fig.~\ref{fig:orientation}d), we notice different orientational responses in the three regions. We quantify these responses using the orientational order parameter \TP{8}{$S_z=\left<(3\cos^2\theta-1)/2\right>$} \cite{Pol2022,Flaig:2024}, where \TP{8}{$\theta$} is the angle between the particle long axis and the penetration direction, with $S_z<0$ indicating preferential transverse alignment. Fig.~\ref{fig:orientation}e shows that $S_z\approx-0.16$ in the bulk region, indicating a weak transverse bias inherited from the initial packing. In the compression-dominated region ahead of the intruder, $S_z$ decreases to approximately $-0.21$ to $-0.28$, indicating enhanced transverse alignment under compression. In the shear-dominated side region, $S_z$ approaches zero and becomes slightly positive, reaching approximately $0.06$, indicating a reorientation toward the penetration direction.
\begin{figure}[htbp]
\centering
\includegraphics[width=\columnwidth]{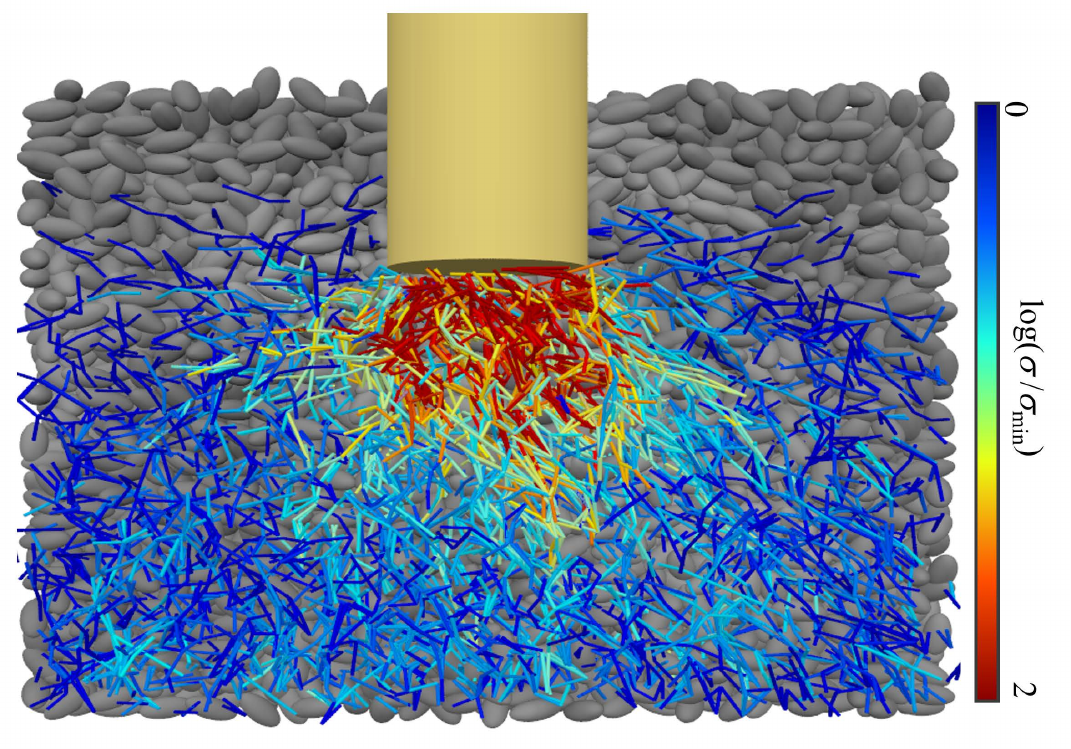}
\caption{Contact network of inter-particle normal forces for $\mathrm{AR}=3$. The color scale represents $\log(\sigma_i/\sigma_{\min})$, where $\sigma_i$ is the minimum principal stress obtained from the coarse-grained local stress tensor within a particle-centered representative volume, and $\sigma_{\min}$ is the minimum value of $\sigma_i$ among all particles.}
\label{fig:forces}
\end{figure}

Particle-scale stresses are strongly localized ahead of the intruder, where highly stressed particles coincide with the dense normal force network (Fig.~\ref{fig:forces}), identifying the compression-dominated front region as the primary load-bearing zone \cite{Peters:2005, Azema:2012, Goldenberg:2002, Lee:2025, Clark:2015}. By contrast, the force network along the intruder side is substantially weaker. This strong front-to-side contrast produces asymmetric contact loading on elongated particles near the transition between the two regions, generating net contact torques that promote particle rotation. Accordingly, the particle angular velocity peaks near this transition (Fig.~\ref{fig:orientation}c), accompanying the reorientation from predominantly transverse alignment ahead of the intruder to a more vertically oriented configuration along its side.

\begin{figure}[htbp]
\centering
\includegraphics[width=\columnwidth]{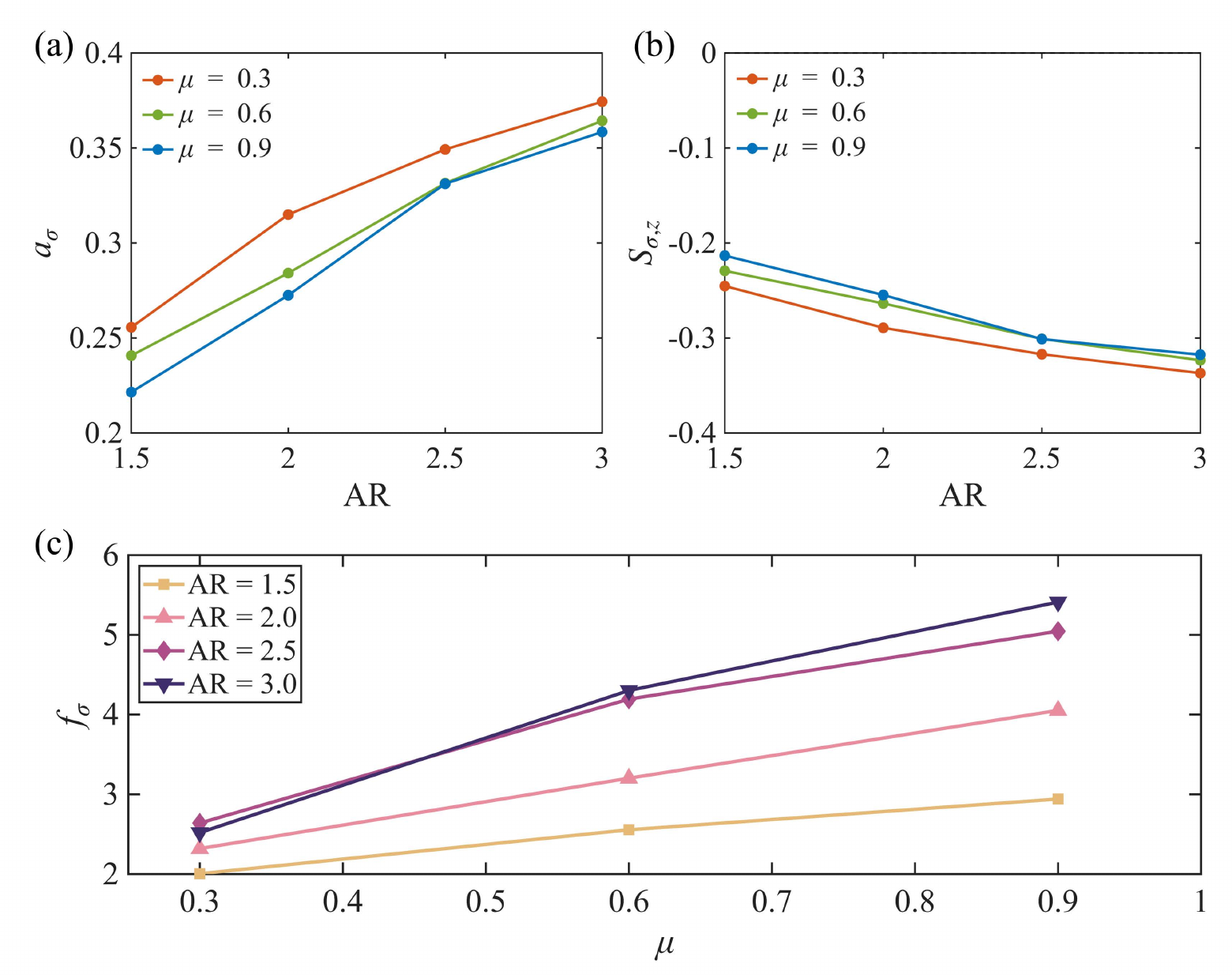}
\caption{\TP{9}{The load-bearing structure is characterized using the high-stress population $\mathrm{SN}$, defined as the $20\%$ of particles with the highest local stresses. a) Stress-weighted orientational anisotropy $a_\sigma$ as a function of $\mathrm{AR}$ for different friction coefficients $\mu$. b) Stress-weighted orientational order parameter $S_{\sigma,z}$ as a function of $\mathrm{AR}$ for different $\mu$. c) Stress amplification factor $f_\sigma$, defined as the ratio of the mean stress in $\mathrm{SN}$ to that in the entire granular assembly, as a function of $\mu$ for different $\mathrm{AR}$.}}
\label{fig:tensor}
\end{figure}

\TP{9}{The high-stress particle population, denoted by $\mathrm{SN}$, comprises the $20\%$ of particles with the highest local stresses. Its orientational organization is characterized by the stress-weighted fabric tensor,}
\begin{equation} 
\hat{T}=\frac{\sum_{i\in \mathrm{SN}} \sigma_i \vec{n}_i\otimes \vec{n}_i}{\sum_{i\in \mathrm{SN}} \sigma_i}\,,
\label{eq:fabric} 
\end{equation}
where $\sigma_i$ is the particle stress and $\vec{n}_i$ is the unit vector along its major axis. Let $\lambda_1\geq\lambda_2\geq\lambda_3$ denote the eigenvalues of $\hat{T}$. The stress-weighted anisotropy $a_\sigma=\lambda_1-\lambda_3$ quantifies the overall degree of orientational anisotropy among the high-stress particles. Small values of $a_\sigma$ correspond to a nearly isotropic distribution of load-bearing orientations, while large values indicate a significant orientational preference. 
\TP{9}{As shown in Fig.~\ref{fig:tensor}a, $a_\sigma$ increases with $\mathrm{AR}$, indicating that particle elongation strengthens the orientational anisotropy in the high-stress population. Its slight decrease with increasing $\mu$ suggests a modest weakening of this anisotropy. While $a_\sigma$ quantifies the overall degree of orientational anisotropy, it does not specify the preferred orientation of the high-stress particles relative to the penetration direction.}

\TP{9}{We therefore introduce the stress-weighted orientational order parameter $S_{\sigma,z}$ to characterize this directional preference as
\begin{equation}
S_{\sigma,z}
=\frac{3\hat{T}_{zz}-1}{2}
=\frac{\sum_{i\in\mathrm{SN}}\sigma_i(3n_{i,z}^{2}-1)}
{2\sum_{i\in\mathrm{SN}}\sigma_i}.
\label{eq:stress_order}
\end{equation}
Here, $n_{i,z}$ is the $z$-component of the particle major-axis unit vector $\vec{n}_i$, with the $z$-axis aligned with the penetration direction. An isotropic distribution of stress-weighted orientations yields $S_{\sigma,z}=0$, while negative and positive values indicating preferential alignment perpendicular and parallel to the penetration direction, respectively. Therefore, $S_{\sigma,z}$ provides directional information complementary to the overall anisotropy quantified by $a_\sigma$. As shown in Fig.~\ref{fig:tensor}b, $S_{\sigma,z}$ remains negative under all simulated conditions and becomes progressively more negative with increasing $\mathrm{AR}$, demonstrating increasingly strong transverse alignment among the high-stress particles. At fixed $\mathrm{AR}$, increasing $\mu$ slightly shifts $S_{\sigma,z}$ toward zero, corresponding to a weakening of this transverse preference.}

The stress amplification factor $f_\sigma=\overline{\sigma}_\mathrm{SN}/\overline{\sigma}_0$ further
characterizes the stress carried by the high-stress population relative to the entire granular assembly,
where $\overline{\sigma}_\mathrm{SN}$ and $\overline{\sigma}_{0}$ denote the mean stresses of the high-stress population and the entire assembly, respectively. Figure~\ref{fig:tensor}c shows that $f_\sigma$ increases both with $\mu$ and $\mathrm{AR}$. The increase with friction is more pronounced for more elongated particles.

\TP{9}{The variations in $a_\sigma$, $S_{\sigma,z}$, and $f_\sigma$
reveal how particle elongation and interparticle friction modify the load-bearing structure. Increasing $\mathrm{AR}$ 
strengthens the orientational anisotropy and transverse alignment of the load-bearing particles, as indicated by the increases in $a_\sigma$ and $|S_{\sigma,z}|$. The simultaneous increase in $f_\sigma$ indicates enhanced stress amplification within the high-stress population.
With increasing $\mathrm{AR}$, high-stress particles exhibit stronger transverse alignment, while their greater length allows them to bridge multiple neighboring particles through spatially separated contacts along their major axes. These contacts mechanically couple neighboring particles and restrict relative translation and rotation, facilitating load transfer through the surrounding particles. This geometric coupling strengthens the load-bearing structure without requiring a comparably large region of particles to move with the intruder, as further corroborated by the observed reduction in stagnant zone volume.
Interparticle friction plays a complementary role. By suppressing contact sliding, increasing $\mu$ allows particles over a broader range of orientations to remain load bearing. Accordingly, $a_\sigma$ and $|S_{\sigma,z}|$ decrease slightly, whereas $f_\sigma$ increases. Particle elongation primarily organizes the load-bearing structure, while friction enhances its stress-bearing capacity.
}

\TP{9}{This interplay between particle shape and friction 
provides a physical basis for the macroscopic scaling $K=K_0(1+b\Pi)$, with $\Pi=(\mathrm{AR}-1)^n\mu^m$, and for the excess resistance above the friction-based reference. 
The increase in penetration resistance with $\mathrm{AR}$ despite the shrinking stagnant zone demonstrates that bulk effective friction alone is insufficient to describe penetration into elongated-particle systems: particle shape-induced orientational organization introduces an additional contribution through anisotropic stress transmission.}

\begin{acknowledgments}
This work was supported by National Natural Science Foundation of China (12602527), Science and Application Research Project for the Tianwen-2 Mission (First Batch of Planetary Exploratolion Project, No. TW2-01-002), China Postdoctoral Science Foundation (No. GZC20251282 and 2025M780854), Fundamental Research Funds for the Central Universities, and Natural Science Basic Research Plan in Shaanxi Province of China (No. 2025JC-QYCX-001).
\end{acknowledgments}
\bibliography{penetration}
\includepdf[pages={{},-}]{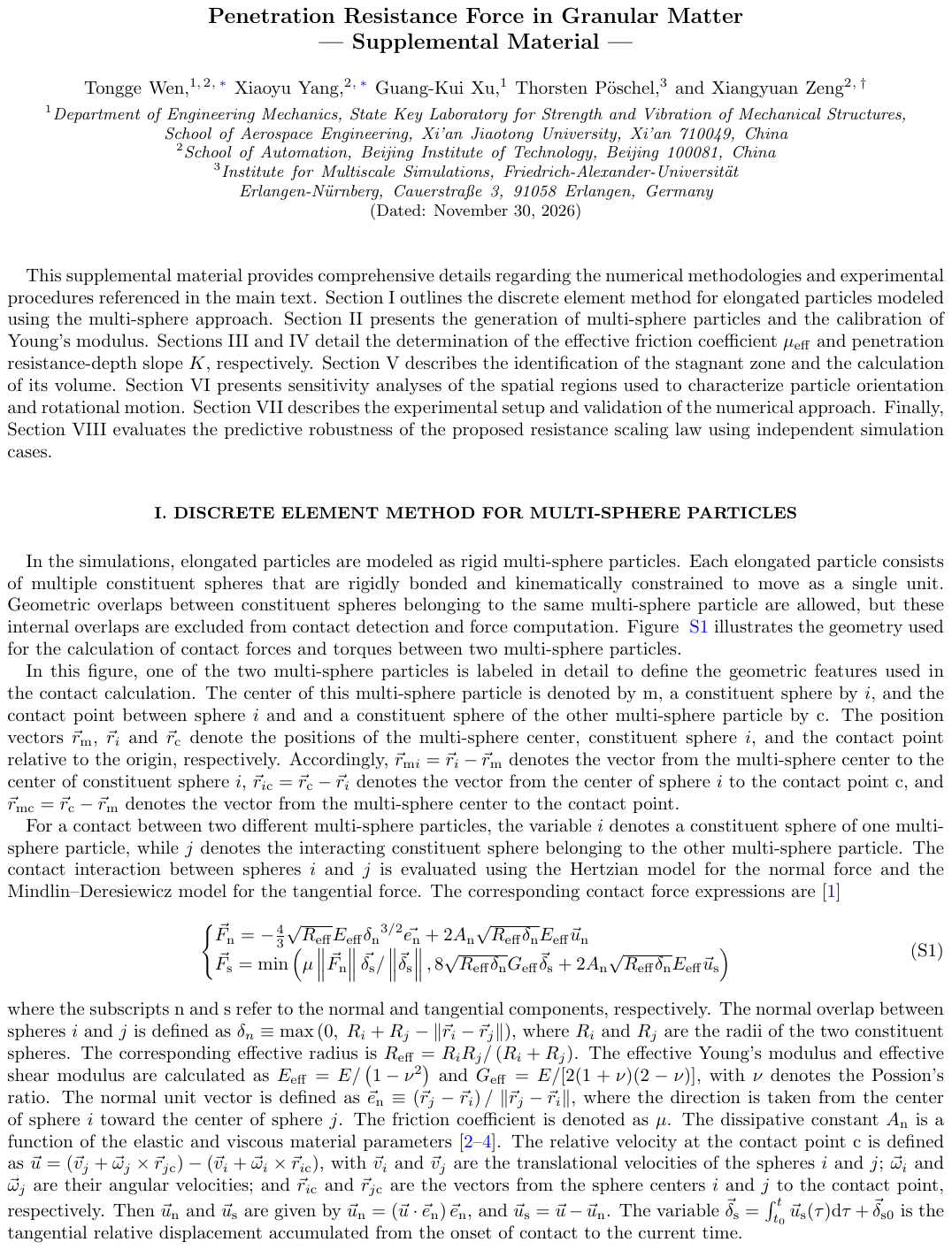}
\AddToHookNext{shipout/before}{\DiscardShipoutBox}
\end{document}